\documentclass[pra,twocolumn,amsmath,amssymb,aps]{revtex4}
\usepackage[utf8]{inputenc}
\usepackage{graphicx}
\usepackage{bm}
\usepackage{subfigure}
\usepackage{amsmath}
\usepackage{color}

\newcommand{\pr}{Phys.\ Rev.\ }

\newcommand{\njp}{New J. Phys.\ }
\newcommand{\etal}{{\em et al.\ }}

\newcommand{\UQ}{School of Mathematics and Physics, University of Queensland, Brisbane, 
QLD 4072, Australia.}
\begin{document}

\title{Negative differential conductivity and quantum statistical effects in a three-site Bose-Hubbard model}

\author{M. K. Olsen and J. F. Corney}
\affiliation{\UQ}
%-----------------------------------------------------------------------
\date{\today}
%------------------------------------------------------------------------

\begin{abstract}

The use of an electron beam to remove ultracold atoms from selected sites in an optical lattice has opened up new opportunities to study transport in quantum systems [R. Labouvie {\it et al.\ }, Phys.\ Rev.\ Lett.\ {\bf 115}, 050601 (2015)]. Inspired by this experimental result,  we examine the effects of number difference, dephasing, and initial quantum statistics on the filling of an initially depleted middle well in the three-well inline Bose-Hubbard model.  
%In a theoretical treatment we are able to examine the effects of number difference, quantum state, and  decoherence separately, and quantify the contribution of each. 
We find that the well-known phenomenon of macroscopic self-trapping is the main contributor to oscillatory negative differential conductivity in our model, with phase diffusion being a secondary effect. However, we find that phase diffusion is required for the production of direct atomic current, with the coherent process showing damped oscillatory currents. We also find that our results are highly dependent on the initial quantum states of the atoms in the system.

\end{abstract}
%******************************************* 

       % check these 

\maketitle

\section{Introduction}
\label{sec:intro}

Recent advances in manipulating ultra-cold bosons in optical traps have opened up new experimental scenarios for experimentally investigating coherent transport phenomena.  Examples of new techniques include  ``potential painting'', which allows for the dynamical formulation of almost arbitrary potentials~\cite{painting}, and the use of an electron beam to empty chosen wells of an optical lattice system~\cite{NDC}. Combined with the well-known techniques of scattering length change via Feshbach resonance~\cite{Feshbach}, such methods allow fine-tuned, dynamical control over nearly all experimental parameters.

For example, Labouvie \etal \cite{NDC} reported on the observation of negative differential conductivity (NDC) of atoms in an optical lattice.  They used the electron beam to remove a proportion of the atoms from a given well and measured the filling rate from the neighbouring wells as a function of chemical potential difference.   The key signature is the decrease in the atomic current as the chemical potential difference is increased, a clear manifestation of non-Ohmic behaviour.  NDC is an unusual phenomenon in electronics because it requires a strongly nonlinear device. In ultracold atom transport, nonlinearity is readily available through atomic collisions. The challenge in a cold-atom implementation of NDC is to obtain a (quasi) steady-state current, which in \cite{NDC} was achieved through dephasing induced  by intrawell multilevel effects.

In this paper, we investigate nonohmic atomic transport in a Bose-Hubbard chain both with and without dephasing.  Although inspired by \cite{NDC}, the model we use is strictly one-dimensional, without the transverse degrees of freedom that dominate the dynamics in that experiment.  There are several reasons for this choice.  First, the simplicity of the Bose-Hubbard model permits us to cleanly investigate the competing effects of nonlinearity and coherence on atom transport.  Second, we are able to explore the role of dephasing per se, without the complications of multiple levels at each site.  And finally, with few modes (1 per site), we enter a regime where quantum effects are important, and which we investigate through the impact of different initial quantum states.

In the electronic context, negative differential conductivity was first described in the transport of electrons in a crystalline superlattice~\cite{Esaki}, where an increase in the electric field resulted in a decrease in electron flow. The phenomenon was subsequently observed in single-molecule junctions~\cite{Perrin}, in suspended metallic single-walled carbon nanotubes~\cite{Pop}, and in graphene transistors~\cite{Britnell}.    %The application of this concept to Bose Einstein condensates (BEC) in an optical lattice is possible when we label the number difference between wells as analogous to voltage and consider that this drives the atomic current. 

In cold atoms systems, NDC and other departures from ohmic behaviour are driven by a number-dependence of the on-site energy or chemical potential, caused by collisional interactions.  In the experimental realisation in  \cite{NDC}, these collisional effects occur in the context of many energy levels per site, which gave rise to dephasing and number-dependent tunnelling.  In the simpler Bose-Hubbard model, the collisional energy differences give rise to macroscopic self-trapping (MST)~\cite{BHJoel,Nemoto,Franzosi,Hines,Albiez}.   
We show that MST by itself is sufficient to cause NDC for initially oscillatory atomic currents.  The addition of sufficient dephasing allows DC current to be observed.

%Our simplified model which reproduces qualitatively the features of the atomic system of Labouvie \etal~\cite{NDC} is a %
We examine the dynamics of the three-well Bose-Hubbard model in which the middle well is initially less occupied than the others. This system without collisional dephasing has previously been analysed by Penna~\cite{CDW}, in what he named the central depleted well regime. Aspects of a similar system have also been analysed by Kordas \etal\cite{Kordas}, who found that phase diffusion could enhance tunneling and damp coherent oscillations in the tunneling. Kordas and others have also analysed the influence of atomic losses on a Bose-Hubbard system, outlining various useful theoretical approaches~\cite{Kordas2}. 

We numerically simulate the quantum dynamics using stochastic equations derived through the truncated Wigner representation.  The simulations quantify the contributions of dephasing, number difference, collisional interaction strength, and initial quantum states to NDC in the three-well model. We find that we can drastically alter the current through the middle well by changing the collisional nonlinearity, which can in principle be done using Feshbach resonance techniques~\cite{Feshbach}. The inclusion of dephasing in the middle well  reduces the currents in some regimes, but does not qualitatively affect the relationship between the maximum current and the number difference. We find that, in contrast, the initial quantum states can have a large effect on the current, demonstrating that the preparation of an experimental system can have a drastic effect on the subsequent dynamics.

\section{Physical model and equations of motion}
\label{sec:model}

Assuming a tight-binding approximation, \cite{BHmodel,Jaksch}, we model the optical lattice system by a three-site Bose-Hubbard chain~\cite{Nemoto,Franzosi,Chiancathermal,BECsplit,splitOC}, with one orbital per site.     We focus on a symmetric distribution of atoms with the middle well initially empty, in analogy to \cite{NDC}, where the electron beam was used to remove atoms from a chosen well. An important feature of the experiment \cite{NDC} was that scattering of atoms amongst the many available radial modes induced an effective phase diffusion in the initially empty well.  We include a controlled level of phase diffusion in our model, which could be implemented by a random variation of the central well depth.  The main effect of phase diffusion is to destroy coherences in the system density matrix, without changing the number distribution~\cite{QNoise}. We note that decay of coherences has been predicted in the Bose-Hubbard model, even without added phase diffusion~\cite{Chiancathermal}.   %, although the decay rate was slower.    % slower than what

Introducing $\hat{a}_{j}$ as the bosonic annihilation operator for atoms in well $j$, we may write the Hamiltonian as
\begin{equation}
{\cal H} =  \hbar\sum_{j=1}^{3}\chi \hat{a}_{j}^{\dag\;2}\hat{a}_{j}^{2}
-\hbar J\left(\hat{a}_{1}^{\dag}\hat{a}_{2}+\hat{a}_{2}^{\dag}\hat{a}_{1}+\hat{a}_{3}^{\dag}\hat{a}_{2}
+\hat{a}_{2}^{\dag}\hat{a}_{3}\right),
\label{eq:Ham3}
\end{equation}
where $J$ is the tunnelling rate between the adjacent wells and $\chi$ is the collisional strength.  The dynamics with phase diffusion are modelled via the master equation
\begin{equation}
\frac{d}{dt}\hat \rho = \frac{1}{i\hbar}\left[{\cal H},\hat \rho \right] + {\cal L}\hat\rho,
\end{equation}
where the Liouvillian superoperator is defined as
\begin{equation}
{\cal L}\hat{\rho} = \Gamma_{2}\left(2\hat{a}_{2}^{\dag}\hat{a}_{2}\hat{\rho}\hat{a}_{2}^{\dag}\hat{a}_{2}-
\hat{a}_{2}^{\dag}\hat{a}_{2}\hat{a}_{2}^{\dag}\hat{a}_{2}\hat{\rho}- \hat{\rho}\hat{a}_{2}^{\dag}\hat{a}_{2}\hat{a}_{2}^{\dag}\hat{a}_{2}\right),
\label{eq:Liouvillian}
\end{equation}
with dephasing rate $\Gamma_{2}$.

Our chosen method for this theoretical investigation is the truncated Wigner approximation~\cite{RobertGraham,Steel}, which %although not exact like the positive-P representation~\cite{Pplus}, 
has been found to be accurate over the time scales we consider here, and which can deal with high nonlinearities. The instances where this method is known to be inaccurate are not relevant to the present work~\cite{Lewis-Swan,kaled}. The integration of the resultant stochastic differential equations is stable and the method does allow us to add more wells with reasonable computational cost. Importantly, the Wigner method can be used to implement the variety of different initial quantum states~\cite{states} that arise in a lattice model.
%This is important because the initial states can have marked effects on the subsequent dynamics and quantum correlations~\cite{Chiancathermal,BECsplit,BH4,BHJPB,BHsteer,BHspread,photostat1,photostat2,photoFock}. We note here that the simulation of coherent states is in principle exact, while that for Fock states is accurate to the order $1/N^{2}$. 
For example, if the system is in the superfluid regime, the appropriate initial states are something close to coherent states. If it is in the Mott insulator regime, or if isolated wells were brought together using recently developed ``potential painting'' techniques~\cite{painting}, the appropriate initial condition would be Fock states.   

Following the usual methods~\cite{DFW}, we map the system master equation for the density operator onto a generalised Fokker-Planck equation for the Wigner distribution. 
%This approach goes beyond the Bogoliubov backreaction method described by Trimborn {\em et al.}~\cite{Trimborn} (see also Davis {\em et al.}~\cite{Thwaite}) in that it includes operator moments to all orders, not imposing any factorisation of these. 
To obtain stable stochastic equations, third-order derivates must be neglected \cite{Steel}. Even with this truncation, the approach goes beyond the Bogoliubov backreaction method \cite{Trimborn,Thwaite}) as it does not impose any factorisation on any higher order moments. %Although third order terms have been modelled using stochastic difference equations~\cite{EPL}, the method is highly unstable, and not at all suitable for our purposes here. 
The resultant stochastic equations in the It\^{o} calculus~\cite{SMCrispin}\footnote{The removal of $\Gamma_{2}$ in the deterministic part of the equations gives the Stratonovich form~\cite{SMCrispin}. Although either form may be used, the Stratonovich form can be integrated using higher order algorithms than the simple Euler method. Care needs to be taken in the design of higher order methods for the integration of It\^{o} equations. The applicability of different methods is covered in the documentation for the open source integration package xmds~\cite{xmds1,xmds2}.}
 are
\begin{eqnarray}
\frac{d\alpha_{1}}{dt} &=& -2i\chi|\alpha_{1}|^{2}\alpha_{1}+iJ\alpha_{2}, \nonumber \\
\frac{d\alpha_{2}}{dt} &=& -\left(\frac{\Gamma_{2}}{2}+2i\chi|\alpha_{2}|^{2} \right)\alpha_{2}+iJ(\alpha_{1}+\alpha_{3})+i\sqrt{\Gamma_{2}}\,\alpha_{2}\eta, \nonumber \\
\frac{d\alpha_{3}}{dt} &=& -2i\chi|\alpha_{3}|^{2}\alpha_{3}+iJ\alpha_{2},
\label{eq:Wigner}
\end{eqnarray}
where $\eta$ is a Gaussian noise with correlations $\overline{\eta}=0$ and $\overline{\eta(t)\eta(t')}=\delta(t-t')$ and the $\alpha_{j}$ are stochastic variables corresponding to the operators $\hat{a}_{j}$. Averages of products of the Wigner variables become approximately equal to the expectation values of symmetrically ordered operator moments in the limit of a large number of stochastic trajectories. We have found previously that the truncated Wigner method gives results for these types of systems that are identical to those of matrix diagonalisation over the times of interest here. The advantage of our method is that the computational complexity is linear in the number of wells, so that more may easily be added.

\section{Results}
\label{sec:results}

We calculate both the average number of atoms in each well, $
N_{j} = \overline{|\alpha_{j}|^{2}}-1/2$,
and the atomic current into the middle well:
%. The current operator~\cite{SantosFilho} is written in terms of the Wigner variables as
\begin{equation}
I_{2} = -i\overline{\left(\alpha_{1}^{\ast}\alpha_{2}-\alpha_{2}^{\ast}\alpha_{1}+\alpha_{3}^{\ast}\alpha_{2}-\alpha_{2}^{\ast}\alpha_{3} \right)}.
\label{eq:currentmean}
\end{equation}
Because the current is generally not steady or even monotonic, the other quantity of interest is the maximum current, $I_{2}^\textrm{max}$, which usually occurs early in the evolution.
To evaluate the current in terms of something analogous to voltage between the middle well and the initially full ones, we define the quantity
\begin{equation}
\Delta\mu\equiv2\chi\left[N_{1}(0)-N_{2}(0)\right],
\label{eq:deltamu}
\end{equation}
which in the thermodynamic limit is proportional to the difference in chemical potential between these wells~\cite{Fisher}. %In the tunneling regime, the definition is not so simple, but because we change $\chi$ in our investigations, we will use this quantity as a parameter.  

%We have initially calculated the quantities of interest for three different values of the collisional nonlinearity $\chi$, $100$ atoms initially in wells $1$ and $3$, and for the middle well with initial populations between $0$ and $50$. %We note here that we are not attempting to duplicate the experimental results obtained by Labouvie {\em et al.}, but are merely investigating the actual quantitative contribution of phase diffusion in the middle well and the initial quantum statistics  to the appearance of NDC. 

\begin{figure}
\begin{center}
\includegraphics[width=0.8\columnwidth]{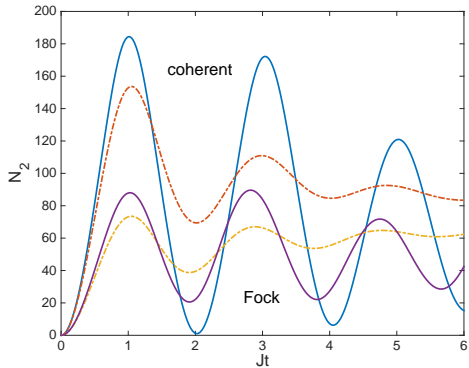}
\end{center}
\caption{(Colour online) The populations in the middle well as a function of dimensionless time, for $J=1$, $\chi=10^{-2}$, and $N_{1}(0)=N_{3}(0)=100$, with $N_{2}(0)=0$. The solid lines represent initial Fock (lower) and coherent (upper) states for $\Gamma_{2}=0$. The dash-dotted lines are the corresponding results for $\Gamma_{2}=1.5J$. All the Wigner results presented in this and subsequent plots are averages of the order of one million stochastic trajectories, with sampling errors  insignificant on the scale of the plots.}
\label{fig:FockConum}
\end{figure}

\begin{figure}
\begin{center}
\includegraphics[width=0.8\columnwidth]{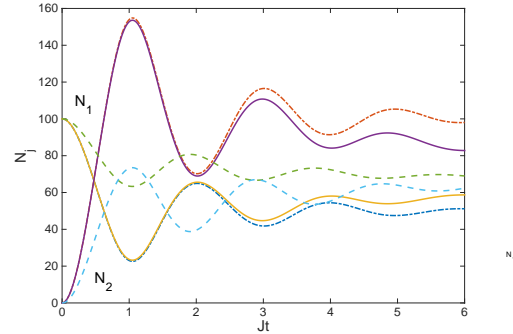}
\end{center}
\caption{(Colour online) The populations in the first and middle wells, for $J=1$, $\chi=10^{-2}$, and $N_{1}(0)=N_{3}(0)=100$, with $N_{2}(0)=0$ and $\Gamma_{2}=1.5$. The solid lines represent initial coherent states, the dashed lines are for initial Fock states and the dash-dotted lines are the results of a classical calculation with diffusion, averaged over $2\times 10^{5}$ trajectories.}
\label{fig:classcompare}
\end{figure}

Figure \ref{fig:FockConum} illustrates the impact of the initial quantum state and phase diffusion on the tunelling dynamics.  We plot the population of the middle well for  $J=1$, $\chi=10^{-2}$, and $N_{1}(0)=N_{3}(0)=100$, with $N_{2}(0)=0$ and $\Gamma_{2}=0$ and $1.5$, for initial Fock and coherent states in the two end wells. We see that the initial quantum statistics have a dramatic effect on the population dynamics whereast the phase diffusion acts mainly to damp out oscillations. Note that $\Gamma_{2}=1.5$ is within the range considered by Labouvie \cite{NDC}. Such quantum statistical effects on the dynamics are similar to those observed in~\cite{Chiancathermal,BH4}   for different Bose-Hubbard configurations, and demonstrate that mean-field approaches are of limited validity. To show this limited validity we have also integrated the classical equations with diffusion, these results being shown in Fig.~\ref{fig:classcompare}. This figure compares the populations in the first and middle wells calculated using initial Fock and coherent states with a completely classical method with added diffusion. The equations look the same as the Wigner equations given  above in Eq.~\ref{eq:Wigner}, but with initial conditions being fixed complex numbers rather than taken from a distribution. What we see is that they are initially almost distinguishable from the coherent state solutions, but diverge from these with time. The maximum currents into well 2 found using this method are virtually identical to those found using coherent states. The difference from the Fock state solutions is more marked, which is to be expected since, while a coherent state is the closest quantum state to a classical state of fixed amplitude and phase, the Fock state is one of the most non-classical states possible.

\begin{figure}
\begin{center}
\includegraphics[width=0.8\columnwidth]{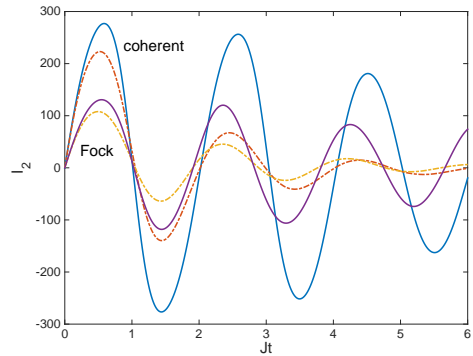}
\end{center}
\caption{(Colour online) The  currents into the middle well  for  the same parameters as Fig.~\ref{fig:FockConum}.  Again, solids lines are without dephasing $\Gamma_{2}=0$ and  dashed lines with dephasing $1.5J$.}
\label{fig:FockCocurrent}
\end{figure}

\begin{figure}
\begin{center}
\includegraphics[width=0.8\columnwidth]{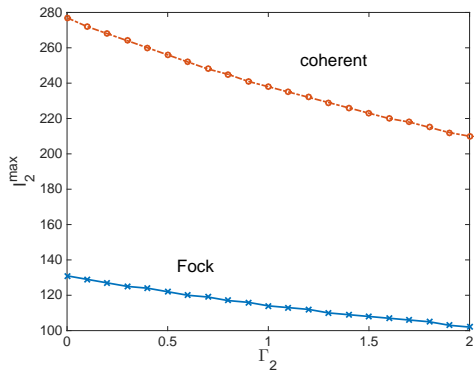}
\end{center}
\caption{(Colour online) Maximum currents into the middle well as a function of $\Gamma_{2}$, for $J=1$, $\chi=10^{-2}$, and $N_{1}(0)=N_{3}(0)=100$, with $N_{2}(0)=0$. The lines are a guide to the eye.}
\label{fig:I2(g2)}
\end{figure}

Figure \ref{fig:FockCocurrent} shows the currents into the middle well for the same parameters used for Fig.~\ref{fig:FockConum}. We see that the maxima of the currents occur for early times, and that these maxima are reduced by the added phase noise (see also Fig.~\ref{fig:I2(g2)}).  Note that a dephasing effect is already present in the unitary evolution, which is a type of quantum collapse phenomena induced by the nonlinearity \cite{BHJoel}.  Thus the currents will damp out even for $\Gamma_{2}=0$ \footnote{Any revival of the oscillations would only occur on timescale vastly longer than what we consider here.}. The additional phase diffusion term causes them to damp out more rapidly, due to the loss of a phase reference between the wells. For all the parameters that we investigate in this article, we find that the first maximum of current is the global maximum, whether we include phase damping or not, and it is this maximum that features in subsequent plots. 

%In Fig.~\ref{fig:I2(g2)} we show the maximum currents as a function of $\Gamma_{2}$, for both Fock and coherent initial states, with the middle well initially empty. What we see is that the maxima decrease as the phase damping is increased, which is consistent with the results of Labouvie \etal 

\begin{figure}
\begin{center}
\includegraphics[width=0.8\columnwidth]{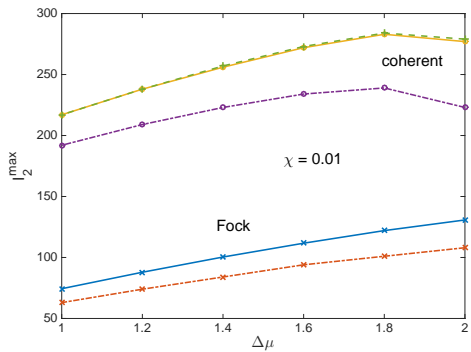}  
\end{center}
\caption{(Colour online) Maximum currents into the middle well as a function of $\Delta\mu=2\chi\left[N_{1}(0)-N_{2}(0)\right]$, for $J=1$, $\chi=0.01$, and $N_{1}(0)=N_{3}(0)=100$, with $N_{2}(0)$ decreasing from $50$ to $0$ along the x-axis. The results for Fock states are represented by the letter x and the circles are for coherent states. The crosses represent the classical values without phase diffusion. The upper plots for each quantum state  are for  $\Gamma_{2}=0$ and the lower plots represent $\Gamma_{2}=1.5J$. The lines are a guide to the eye.}
\label{fig:current01}
\end{figure}

To determine the impact of quantum effects on the conductivity, we calculate the maxima of the currents as a function of  $\Delta\mu$  for different initial quantum states, both with and without phase diffusion.  Fig.~\ref{fig:current01} shows the maximum currents as a function of $\Delta\mu$, for $\chi=0.01$, with $N_{2}(0)$ varying from $50$ to $0$ along the horizontal axis. For the initial Fock states, the maximum current is increases almost linearly over this range, whereas for the initial coherent state, clear departures from Ohmic behaviour set in by $N_{2}(0)=40$ $(\Delta\mu=1.8)$.  In particular the current for coherent state starts to decrease with further increases in chemical potential difference, which we may regard as a type of negative differential conductivity caused by  macroscopic self-trapping.  Note that these features are qualitatively the same whether the additional dephasing is considered or not. %In fact, it happens even for the classical prediction, which is almost identical to the truncated Wigner results with initial coherent states for these parameters.

\begin{figure}
\begin{center}
\includegraphics[width=0.8\columnwidth]{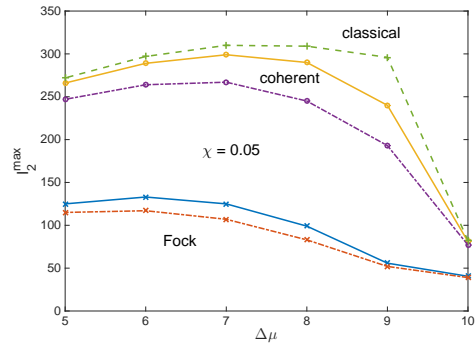}    
\end{center}
\caption{(Colour online) The  maximum currents into the middle well as a function of $\Delta\mu$, for $J=1$, $\chi=0.05$, and $N_{1}(0)=N_{3}(0)=100$, with $N_{2}(0)$ decreasing from $50$ to $0$ along the x-axis. Plot styles are as in Fig.~\ref{fig:current01}.}
\label{fig:current05}
\end{figure}

We can increase the NDC effect markedly by increasing the nonlinearity to $\chi=0.05$, as shown in Fig.~\ref{fig:current05}. In this case we again see marked differences depending on the initial quantum states, with the addition of finite $\Gamma_{2}$ having only a quantitative effect. We see NDC for values of $\Delta\mu >6$ for Fock states, and $\Delta\mu>7$ for coherent states. In this case we see that both quantum solutions show NDC for slightly lower values of $\Delta\mu$ than found in the mean-field prediction. This is explained by the appearance of macroscopic self-trapping at slighty different values of $\Delta\mu$ for the quantum solutions. We define the onset of macroscopic self-trapping as occurring when the population in the middle well always remains less than the populations in the two outside wells.  A numerical investigation of the classical system shows that the MST onset happens at a value of $\Delta\mu=7.82$. For a quantum system, we do not expect such a sharp transition, since unlike the classical case, it does not have a precisely defined atom number and phase.  Coherent states are the closest to classical states, but have a Poissonian number distribution, the lower values supporting full population oscillations. The initial Fock states give a definite number initially, but it is still the case that during the evolution, neither the relative number or phase is precisely defined.

\begin{figure}
\begin{center}
\includegraphics[width=0.8\columnwidth]{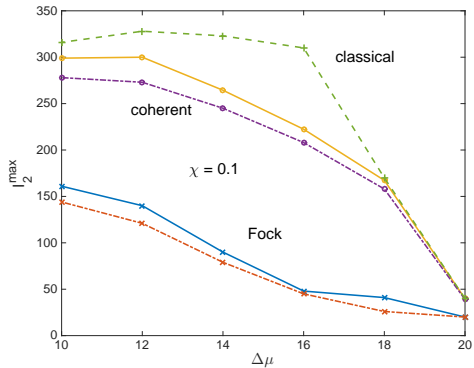}  
\end{center}
\caption{(Colour online) The  maximum currents into the middle well as a function of $\Delta\mu$, for $J=1$, $\chi=0.1$, and $N_{1}(0)=N_{3}(0)=100$, with $N_{2}(0)$ decreasing from $50$ to $0$ along the x-axis. Plot styles are as in Fig.~\ref{fig:current01}.}
\label{fig:current1}
\end{figure}

In Fig.~\ref{fig:current1}, with a further increase in interaction strength ($\chi=0.1$), the NDC is more marked, especially for the initial Fock states.  The whole range of $\Delta\mu$ considered in this graph is within the classical MST regime: although a smaller fraction of atoms is transferred, the tunnelling occurs at a more rapid rate, leading to a larger maximum current  than in the previous two figures

\section{Direct current tunnelling}
\label{sec:mais}

One of the interesting features reported by Labouvie \etal was direct current atom tunneling. Although their experiment was more complicated than our simple three-mode model, having more wells and including radially excited levels in the larger traps, we find that we are able to reproduce DC current in our system. We find that this happens deep in the MST regime, where only small population oscillations are seen classically. As an example, we have chosen $\chi=0.01$, with 700 atoms in each of the outside wells. This gives $\Delta\mu=14$ and the results are similar for the other two values of $\chi$ used above, as long as the initial numbers are changed so that $\Delta\mu$ remains constant. 
This DC tunneling is completely caused by the added phase diffusion, with the rate also being dependent on the initial quantum states.

\begin{figure}
\begin{center}
\includegraphics[width=0.8\columnwidth]{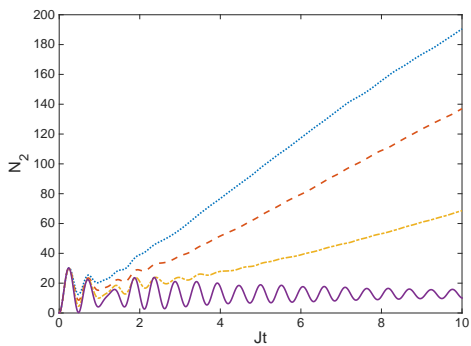}  
\end{center}
\caption{(Colour online) The populations of the middle well for initial Fock states with larger number.  Here $J=1$, $\chi=0.01$, and $N_{1}(0)=N_{3}(0)=700$, for different values of $\Gamma_{2}$. The solid line is $\Gamma_{2}=0$, the dash-dotted line is $\Gamma_{2}=0.5$, the dashed line is $\Gamma_{2}=1$, and the dotted line is $\Gamma_{2}=1.5$.}
\label{fig:N2F700}
\end{figure}

\begin{figure}
\begin{center}
\includegraphics[width=0.8\columnwidth]{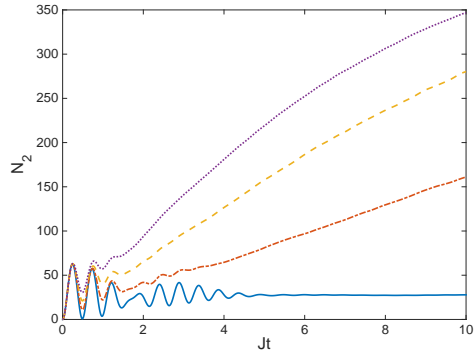}  
\end{center}
\caption{(Colour online) The populations of the middle well for initial coherent states with large average number. Parameters as in Fig,\ \ref{fig:N2F700}.}
\label{fig:N2C700}
\end{figure}

In Fig.~\ref{fig:N2F700} we show the results for initial Fock states. Without added phase diffusion, we see slowly damped regular oscillations, with only a small number of atoms ever entering the middle well. In contrast, with finite $\Gamma_{2}$, the atom number grows almost linearly after small initial oscillations. Figure \ref{fig:N2C700} shows the results for initial coherent states. We see larger magnitude initial oscillations, with the population settling to a higher equilibrium value for $\Gamma=0$ than in the Fock state case. The populations for $\Gamma\neq 0$ are higher than for the case of initial Fock states, but again there is an almost linear growth in middle-well population. Obviously this positive population transfer cannot last forever, as the outside wells will run out of atoms to transfer. Running our simulations for longer times indicate that the populations in the three wells become equal, with approximately $470$ atoms in each well. At this stage, all the coherences have decayed to zero and the tunneling current stops.

\section{Conclusions}
\label{sec:conclusions}

In conclusion, we have investigated atomic transport  in a three-well Bose-Hubbard mode, revealing a type of negative differential conductivity driven by macroscopic self trapping. With moderate atom numbers in each of the outside wells, we find oscillatory populations, with the maximum of the tunneling current always on the first oscillation. The major quantitative influence on the size of these currents is the initial quantum states of the atomic modes in the outside wells, with marked differences between Fock and coherent states. Given recent experimental advances in the preparation of ultra-cold atomic systems, these predictions should be amenable to experimental investigation in the near future.

The addition of phase diffusion serves to lower the current maxima in this oscillatory regime, but does not change the differential atomic conductivity from positive to negative. We find that, with added phase diffusion, the coherent state results are very similar to those found in a classical diffusive model. Overall, the extra dephasing makes a quantitative, rather than qualitative, contribution to the current.  Moreover,  the DC tunnelling regime induced by dephasing only occurs for sufficiently large initial number difference.

\section*{Acknowledgments}

This research was supported by the Australian Research Council under the Future Fellowships Program (Grant ID: FT100100515).

\end{document}